\documentclass[a4paper,prb,twocolumn,showpacs]{revtex4}
\usepackage[dvips]{graphicx}
\usepackage{amsmath}
\usepackage{array}
\usepackage{bm}
\usepackage{fancyhdr}
\usepackage{makeidx}
\usepackage{epsfig}
\usepackage{amsfonts}
\usepackage{amssymb}
\usepackage{color}

\begin{document}

\title{Time-dependent electron transport through a strongly correlated quantum dot: 
multiple-probe open boundary conditions approach}
\author{A. Pertsova, M. Stamenova and S. Sanvito}
\affiliation{School of Physics and CRANN, Trinity College Dublin, Dublin 2, Ireland}
\date{\today}

\begin{abstract}
We present a time-dependent study of electron transport through a strongly correlated quantum dot. The time-dependent 
current is obtained with the multiple-probe battery method, while adiabatic lattice density functional theory in the Bethe ansatz 
local-density approximation to the Hubbard model describes the dot electronic structure. We show that for a certain range of  
voltages the quantum dot can be driven into a dynamical state characterized by regular current oscillations. This is a manifestation 
of a recently proposed dynamical picture of Coulomb blockade. Furthermore, we investigate how the various approximations to 
the electron-electron interaction affect the line-shapes of the Coulomb peaks and the \textit{I-V} characteristics. We show that the 
presence of the derivative discontinuity in the approximate exchange-correlation potential leads to significantly different results 
compared to those obtained at the simpler Hartree level of description. In particular, a negative differential conductance (NDC) in 
the \textit{I-V} characteristics is observed at large bias voltages and large Coulomb interaction strengths. We demonstrate that such 
NDC originates from the combined effect of electron-electron interaction in the dot and the finite bandwidth of the electrodes. 
\end{abstract}
 
\maketitle

\section{Introduction}\label{intro}

Electron transport through nanoscale devices is a diverse subject, which is currently the focus of extensive experimental and 
theoretical research. The fuel of such interest is the expectation that nanoscale objects, such as quantum dots~\cite{Ashoori} 
and even single molecules,~\cite{Ratner} are to become active components in novel electronic devices, which potentially offer 
unique advantages over existing technologies.~\cite{molelec2} At the fundamental level, the physics of such reduced-dimensional 
systems is dominated by quantum effects. Among them are electron correlations, which strongly affect the electron transport at 
this level of confinement, giving rise to prototypical quantum phenomena, such as Coulomb blockade~\cite{cb2,cb3} and the 
Kondo effect.~\cite{kondo1,kondo2,kondo5}

While the Landauer formula is the solution to the non-interacting quantum transport problem,~\cite{Landauer} the interacting case continues to be challenging to the theory. The latter is typically approached with the non-equilibrium Green's function (NEGF) formalism,~\cite{Haug} which allows, in principle, the derivation of an interacting many-body Landauer-type formula for the steady-state current in the case where interaction is limited to a finite region in space.~\cite{Meir} In practice, for the majority of the state-of-the-art \textit{ab initio} transport calculations and numerical algorithms,~\cite{Stefano,Ryndyk,smeagol1,smeagol2} the method of choice for the electronic structure description is the density functional theory (DFT). However, typical steady-state DFT+NEGF transport schemes have a range of limitations, both conceptual and technical.~\cite{Kurth_method} 

At the fundamental level it has been recently demonstrated, at least for the case of a single Anderson impurity model, that the linear response 
conductance calculated from the Kohn-Sham levels for the exact exchange-correlation (XC) functional reproduces closely that computed with 
many-body approaches.~\cite{Evers,Stafford} If the same holds true for {\it ab initio} DFT, then the DFT+NEGF scheme will provide a complete solution 
for the zero-bias limit. Still, on the practical side, the commonly used approximations to the XC functional, lacking the so-important derivative 
discontinuity,~\cite{Perdew} fail to capture essential physics for the transport in molecular junctions, qualitatively mispredicting the conduction 
regime.~\cite{Toher,Toher2} Different is the situation at finite bias, where, let alone the implementation, conceptual concerns reflect on the 
very applicability of a ground-state electronic structure theory to an intrinsically non-equilibrium problem especially if electron correlations 
are significant.~\cite{Koentopp,Vignale} 

One strategy to avoid some of the shortcomings of using equilibrium DFT has been sought in its natural extension, time-dependent 
(TD) DFT,~\cite{tddft} with practical schemes for TD transport having been developed.~\cite{Kurth_method} In general, real-time 
TD schemes for quantum transport can be roughly divided into two categories based on their assumption for the initial conditions. In one case 
the electrodes are prepared in equilibrium with the poles of a battery, but not yet connected to the nanoscopic device. The current then
starts to flow when the connection is made. In the other the system electrodes+device is initially at equilibrium and subsequently 
an electric field is applied to the electrodes. The former assumption, where two initial electrochemical potentials are well defined, is more in 
the spirit of the Landauer transport picture. The latter is instead more DFT-friendly, as the starting point is the ground state of the 
system.~\cite{Kurth_method} 

There has been evidence that these two TD transport variants agree in the non-interacting case, i.e. they lead to the same history-independent 
steady-state current.~\cite{Cini,Stefanucci} More recently, the latter variant combined with the TDDFT, further equipped with a novel XC functional 
carrying the physical derivative discontinuity, has been applied to study the transport through a quantum dot in the Coulomb blockade (CB) regime 
by Kurth \textit{et al.} in Ref.~[\onlinecite{Kurth}]. In particular that work has put forward an important novel description of CB as a dynamical process 
with rapidly oscillating local currents, inaccessible by conventional steady-state transport models.

In this work we adopt another recently proposed TD transport scheme, the so-called, \textit{multiple-probe battery} (MPB) method,~\cite{Todorov1,Todorov2} 
to study electron transport through a strongly correlated quantum dot. The MPB scheme was first proposed in the context of correlated electron-ion 
dynamics and was applied to a wide range of problems, such as current-induced heating in atomic wires.~\cite{Todorov1, Todorov3} This method 
belongs to the first of the fore-mentioned categories and enables the realization of an external battery within the finite system of electrodes+device. The 
external bias is introduced through the difference in the electrochemical potentials of the set of reservoirs, or \textit{probes}, attached individually to each atom 
in a pair of large but finite metallic electrodes (leads). The scheme is very tractable computationally and has the control knobs to be an arbitrarily close 
approximation to the non-interacting Landauer transport in the long-time dc limit.

The MPB time-propagation scheme is based on the integration of the Liouville-von Neumann equation of motion for the reduced density matrix of the system, 
in which the open boundaries are described explicitly by a source and a drain term. For the TD Hamiltonian of the quantum dot, entering the equation of motion, 
we adopt the description used by Kurth \textit{et al.}.~\cite{Kurth} This is based on the adiabatic Bethe ansatz local-density approximation~\cite{balda1} (adiabatic 
BALDA, or ABALDA) to the XC functional, which exhibits a derivative discontinuity at half-filling.

By investigating the real-time evolution of the current through the quantum dot, we find an agreement with Ref.~[\onlinecite{Kurth}], i.e. for a certain set of 
parameters the system does not reach a steady state but rather remains in a dynamical state, characterized by oscillations in the current. Furthermore, we try 
to interpret the TD results in terms of the more familiar steady-state picture of transport. In particular, we construct the current-voltage, \textit{I-V}, characteristics 
of the quantum dot from the long-time average of the current and the voltage obtained from the TD simulations. This is done for a wide range of parameters, even in 
the cases when a steady state is not achieved. Importantly, we observe a drop of the current as a function of the source-drain voltage and, as a consequence, 
a negative differential conductance (NDC) above a critical bias voltage. We demonstrate that such an effect is not possible if the derivative-discontinuity is not
included in the one-particle potential. 

This is particularly interesting in view of some recent contrasting results. On the one hand a number of studies, based on several distinct many-body 
approaches,~\cite{Nishino0,Doyon,Boulat} attribute the NDC mainly  to electron-electron interaction. On the other hand, it has been demonstrated  by 
B\^{a}ldea and K\"oppel \cite{Baldea} that in the case of an exactly solvable model for a non-interacting dot within the steady-state formalism, the finite bandwidth 
of the electrodes can alone lead to pronounced NDC for a wide range of parameters. Here we find a numerical proof that this result can be generalized to the 
interacting case and time-dependent transport. Our calculations suggest, however, that for the system considered here, the NDC is due to a combination of two 
effects, namely electron-electron interaction on the dot {\it and} the finite bandwidth of the electrodes.

Our paper is organized as follows. In the next section we introduce the model system and our theoretical framework, i.e. the Hamiltonian 
and the computational scheme for MPB quantum transport. In the first part of Section~\ref{results} the \textit{I-V} characteristics of a 
non-interacting quantum dot calculated by using the TD-MPB method is compared to analytic NEGF results. We then discuss the finite 
electrode bandwidth as a source of NDC. In the second  part of Section~\ref{results}, we present the TD results for a strongly correlated dot in 
the CB regime. Finally, we propose an explanation for the observed NDC in the \textit{I-V} characteristics.

\section{Methods}\label{theo}

\begin{figure}[ht]
\begin{center}\includegraphics[width=0.97\linewidth,clip=true]{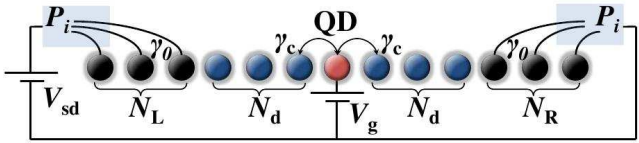}\end{center}
\caption{(Color online) Schematic of the model system considered in this work: the central region consists of a quantum dot (QD) surrounded 
by two $N_\mathrm{d}$-site long atomic chains, which in turns are attached to two one-dimensional leads comprising respectively $N_\mathrm{L}$ 
and $N_\mathrm{R}$ sites. Here $\gamma_0$ is the hopping integral in the leads and in the two chains, and $\gamma_\mathrm{c}$ is the 
lead to dot hopping. $V_\mathrm{g}$ denotes the gate voltage, acting locally on the dot, and $V_\mathrm{sd}$ is the source-drain voltage 
applied across the entire system.}
\label{qd}
\end{figure}

The model system considered in this work is presented in Fig.~\ref{qd}. This consists of a central region, which contains 
the quantum dot surrounded by two $N_{\mathrm{d}}$-site long atomic chains at both sides, and two one-dimensional finite 
leads, each counting $N_\mathrm{L(R)}$ atoms. The physics of the quantum dot connected to two leads is described by the 
Anderson impurity model.~\cite{Gruner,Meir} The Hamiltonian of the total system thus reads
\begin{equation}
 \hat{H}_\mathrm{S} = \sum_{\scriptsize \alpha=\mathrm{L,R}}\hat{H}_{\alpha}+\hat{H}_\mathrm{T}+\hat{H}_\mathrm{QD}\:.
 \label{eq:1}
\end{equation}
Here the first term is the nearest-neighbors single-orbital tight-binding (TB) Hamiltonian describing respectively the left-hand side ($\alpha$=L) 
and right-hand side ($\alpha$=R) lead. This is written as
\begin{equation}
\hat{H}_{\alpha}=\sum_{\scriptsize i,\sigma} \varepsilon_{i\alpha}\, \hat{c}^{\sigma\dagger}_{i\alpha} \,\hat{c}^{\sigma}_{i\alpha}+
\sum_{\scriptsize i,\sigma} \gamma_0\left(\,\hat{c}^{\sigma\dagger}_{i\alpha} \,\hat{c}^{\sigma}_{i+1\alpha} + h.c.\right)\:,
\label{eq:1a}
\end{equation}
where $\varepsilon_{i\alpha}$ are the on-site energies and $\gamma_0$ is the hopping integral; 
$\hat{c}_{i\alpha}^{\sigma\dagger}(\hat{c}_{i\alpha}^{\sigma})$ is the creation (annihilation) operator for an electron with spin
 $\sigma$ ($\sigma$=$\uparrow,\downarrow$) at the atomic site $i$ of the  lead $\alpha$ (the index $i=1,..,N_{\alpha}$ runs from left 
to right for $\alpha$=R and from right to left for $\alpha$=L). Note that two atomic chains on each side of the quantum dot are also 
described by a TB model with the hopping integral $\gamma_0$ and therefore they are included in the Hamiltonian of the leads.

The second term in Eq.~(\ref{eq:1}) describes the tunneling between the quantum dot and the two adjacent sites and it is given by
\begin{equation}
 \hat{H}_\mathrm{T}=\sum_{\scriptsize \sigma} \gamma_\mathrm{c}\left(\,\hat{c}^{\sigma\dagger}_{0} \,\hat{c}^{\sigma}_{1L} + 
\, \hat{c}^{\sigma\dagger}_{0} \,\hat{c}^{\sigma}_{1R} + h.c.\right), 
\label{eq:1b}
\end{equation}
where $\hat{c}_{0}^{\sigma\dagger}(\hat{c}_{0}^{\sigma})$ is the creation (annihilation) operator for an electron with spin $\sigma$ on the dot
 and $\gamma_\mathrm{c}$ is the hopping integral between the dot and site $i$=$1$ in the lead $\alpha$.

Finally, the Hamiltonian of the quantum dot reads
\begin{equation}
 \hat{H}_\mathrm{QD}=\sum_{\scriptsize \sigma} V_\mathrm{g}\,\hat{n}_{0}^{\sigma}+U\,\hat{n}_{0}^{\uparrow}\hat{n}_{0}^{\downarrow}, 
\label{eq:1c}
\end{equation}
where $V_\mathrm{g}$ is the on-site energy of the dot, which acts as a local gate voltage; $U$ ($U\ge 0$) is the charging energy, 
which expresses the strength of the Coulomb repulsion on the dot; $\hat{n}_{0}^{\sigma}$=$\hat{c}^{\sigma\dagger}_{0}\,\hat{c}^{\sigma}_{0}$ 
is the site-occupation operator.

Within the lattice DFT framework~\cite{LDFT} the many-body Hamiltonian in Eq.~(\ref{eq:1c})  is mapped onto an effective 
single-particle Kohn-Sham Hamiltonian which, in the local density approximation, reads
\begin{equation}
 \hat{H}_\mathrm{0}=\sum_{\scriptsize \sigma} v_\mathrm{KS}\left[ n_{0}\right] \,\hat{n}_{0}^{\sigma}.  
\label{eq:2}
\end{equation}
Here $n_{0}$ is the charge density of the dot and  $v_\mathrm{KS}$ is the effective Kohn-Sham potential, which can be 
written as a sum of three terms
\begin{equation}
 v_\mathrm{KS}\left[n_{0}\right] =V_\mathrm{g}+\frac{n_{0}}{2}U+v_\mathrm{XC}\left[ n_{0}\right] . 
\label{eq:3}
\end{equation}
The second and third terms are respectively the Hartree and the XC potential. The latter is approximated  
by a modified BALDA potential, specifically tailored to a nonuniform configuration with a weakly coupled dot
(we refer to Ref.~[\onlinecite{Kurth}] for the exact expression and the parametrization). 

Notably, such $v_\mathrm{XC}$ exhibits a derivative discontinuity at $n_0$=$1$, i.e. at the phase transition of
the 1D Hubbard model. In practice, however, we use a continuous approximation to the BALDA potential~\cite{Kurth} 
where the true discontinuity, expressed through a Heaviside step function $\theta(n_0)$, is replaced by a function 
$f(n_0)=1/(e^{(n_0-1)/a}+1)$ with $a$ being a smoothing parameter. We use $a=10^{-7}$, which guarantees a very 
sharp slope at $n_0=1$. In our simulations we consider three levels of description: \textit{(i}) $U=0$, or non-interacting 
case, for which the effective potential of the dot is simply given by $v_\mathrm{KS}$=$V_\mathrm{g}$, 
\textit{(ii})  $v_\mathrm{XC}\rightarrow 0$, or the Hartree approximation, where the potential on the dot is    
$v_\mathrm{H}$=$V_\mathrm{g}+U\,n_{0}/2$; and \textit{(iii}) the full discontinuous effective potential, given by 
Eq.~(\ref{eq:3}), which we refer to as  $v_\mathrm{KS}$ for clarity.

In order to introduce the time-dependence in the Hamiltonian of the quantum dot, we use the adiabatic approximation, 
where $v_\mathrm{0}$ is assumed to depend on time only through the instantaneous charge density of the dot 
\begin{equation}
 v_\mathrm{KS}(t)=v_\mathrm{KS}[n_0(t)]\:.\label{eq:3a}
\end{equation}

The question of the applicability of such adiabatic local approximation to the description of non-equilibrium transport in 
strongly correlated systems has been addressed in recent two works respectively by Uimonen \textit{et al.}~\cite{Uimonen} and Khorsavi 
\textit{et al.}~\cite{Khorsavi} In particular, a comparative study between the TDDFT approach with ABALDA (TDDFT+ABALDA) 
and the many-body perturbation theory, applied to out-of-equilibrium Anderson impurity model, has been carried out in 
Ref.~[\onlinecite{Uimonen}]. The results obtained with both approaches have been tested against numerically exact results 
produced by time-dependent density matrix renormalization group theory. It was found that, in general, the TDDFT+ABALDA 
approach is in good qualitative agreement with many-body perturbation theory over a wide range of parameters. However, in many 
cases it overestimates the steady-state currents. This problem was linked to the shortcomings of the local approximation to the XC 
functional and, in particular, to the absence of electron correlations inside the electrodes. Moreover, it was demonstrated in 
Ref.~[\onlinecite{Khorsavi}] that the inclusion of dynamical correlations, or memory effects, might eliminate the  multistability in the 
density and the current, which can be found  within the TDDFT+ABALDA approach. These are strong indications that more advanced 
non-local, both in space and time, approximations to the XC functional are required. However, as was demonstrated in 
Ref.~[\onlinecite{Kurth}] and as it will be shown in this paper, the ABALDA already provides valuable insights into time-dependent 
transport in strongly correlated systems. 

We now discuss, following the work of Todorov and co-workers,~\cite{Todorov1,Todorov2} how the open boundary conditions are 
introduced in the MPB setup. In the MPB method, each atom $i$ of the leads (with the exception of the $N_\mathrm{d}$ atoms at 
both sides of the quantum dot) is connected to an external probe $P_i$ (see Fig.~\ref{qd}). All the probes attached to the sites in the left 
(right) lead are kept at the electrochemical potential $\mu_\mathrm{L}$ ($\mu_\mathrm{R}$) and are occupied according to the 
Fermi-Dirac distribution $f_\mathrm{L}$ ($f_\mathrm{R})$. The source-drain voltage $V_\mathrm{sd}$ is introduced as 
$V_\mathrm{sd}=\mu_\mathrm{L}-\mu_\mathrm{R}$ (here $V_\mathrm{sd}$ is in units of eV). For symmetrically applied 
bias $\mu_\mathrm{L}=\varepsilon_\mathrm{F}+V_\mathrm{sd}/2$ and $\mu_\mathrm{R}=\varepsilon_\mathrm{F}-V_\mathrm{sd}/2$, 
where $\varepsilon_\mathrm{F}$ is the Fermi level of the electrodes (assumed identical). The time-dependent equation of
motion for the density matrix of the system coupled to the probes reads
\begin{eqnarray}
 i\hbar\,\dot{\hat{\rho}}_\mathrm{S}(t) & = & \left[\hat{H}_\mathrm{S}(t), \hat{\rho}_\mathrm{S}(t)\right]+\hat{\Sigma}^{+}\,\hat{\rho}_\mathrm{S}(t)
-\hat{\rho}_\mathrm{S}(t)\,
\hat{\Sigma}^{-}+\label{eq:4}\\
& + & \int_{-\infty}^{\infty} \left[  \hat{\Sigma}^{<}(E)\,\hat{G}^{-}_\mathrm{S}(E)-\hat{G}^{+}_\mathrm{S}(E)\,\hat{\Sigma}^{<}(E) \right]  dE\:. \nonumber
\end{eqnarray}
The last two terms on the right-hand side are extraction (drain) and injection (source) terms, respectively; $\hat{G}^{+}$ ($\hat{G}^{-}$) 
is the retarded (advanced) Green's function of the system and it is given by
\begin{equation}
 \hat{G}^{\pm}=\left( E\,\hat{I}_\mathrm{S} - \hat{H}_\mathrm{S_0} -\hat{\Sigma}^{\pm} \pm i\,\hat{I}_\mathrm{S}\,\Delta\right)^{-1}\:,
\label{eq:5}
\end{equation}
where $\hat{H}_\mathrm{S_0}= \sum_{\scriptsize \alpha=\mathrm{L,R}}\hat{H}_{\alpha}+\hat{H}_\mathrm{T}+
\sum_{\scriptsize \sigma} V_\mathrm{g}\,\hat{n}_{0}^{\sigma}$ is the time-independent part of $\hat{H}_\mathrm{S}(t)$ and $\Delta$ 
is a dephasing factor (see later for an exact definition). The self-energies due to the presence of the external probes and the in-scattering 
self-energy are written as  
\begin{eqnarray}
 \hat{\Sigma}^{\pm} & = & \mp i\, \frac{\Gamma}{2}\,\hat{I}_\mathrm{L} \mp i\, \frac{\Gamma}{2}\,\hat{I}_\mathrm{R}\:,
\label{eq:6}\\
\hat{\Sigma}^{<} & = & \frac{\Gamma}{2\pi}\,f_\mathrm{L}(E)\,\hat{I}_\mathrm{L} + \frac{\Gamma}{2\pi}\,f_\mathrm{R}(E)\,\hat{I}_\mathrm{R}\:,
\label{eq:7}
\end{eqnarray}
with the broadening $\Gamma$ defined as $\Gamma=2\pi\gamma_{P}^{2}d$, where $\gamma_P$ is the coupling to the probes, 
assumed to be identical for all sites in the leads, and $d$ is an energy-independent constant, which represents the surface density 
of states of the probes within the wide-band limit; $\hat{I}_\mathrm{M}$ is the identity operator in region M (M=L, R, S).

Equation (\ref{eq:4}) is derived from a general Liouville-von Neumann equation for the total density matrix of the system and the 
probes combined. It incorporates two main approximations: \textit{(i)} the wide-band limit in the probes and 
\textit{(ii)} the decoherence in the injection process, introduced through the relaxation time $\tau_{\Delta}$, with $\Delta=\hbar/\tau_{\Delta}$ 
[see Eq.~(\ref{eq:5})]. The second approximation essentially decouples, over the time interval $\tau_\Delta$, the injection of electrons 
from the probes into the leads and their subsequent scattering from the time-dependent potential inside the central region, provided that 
the latter is long enough. In other words the dephasing factor imposes a restriction on the size of the central region ($2 N_\mathrm{d}+1$ sites). 
Therefore the inclusion of $N_\mathrm{d}$ buffer sites on both sides of the dot is essential within the time-dependent formalism. 

The value of $\Delta$ is determined  in such way that the distance traveled by the electrons during the time interval $\tau_\Delta$ is smaller 
than the distance between the electrodes and the interior of the central region, i.e. the quantum dot. This condition can be written as 
$v_e\tau_{\Delta}<N_\mathrm{d}a$, where $v_e$ is the electron group velocity and $a$ the lattice constant ($a=1$). 
In practical terms, the introduction of the dephasing factor allows one to write down the injection term, which is in general non-local in 
time, in a rather simple time-independent form [see Eq.~(\ref{eq:4})]. This, however, also introduces an additional broadening, proportional 
to $\Delta$, in the steady-state \textit{I-V} characteristics, which is absent in the standard static NEGF formalism. We note that in the 
steady-state MPB formalism, the NEGF result is recovered in the limit of infinitely long leads and weak lead-probe coupling.~\cite{Todorov2}

In order to investigate the open-boundary electron dynamics in the time domain, Eq.~(\ref{eq:4}) is numerically-integrated using the 
fourth-order Runge-Kutta (RK4) algorithm.~\cite{RK4} As initial condition, we use the density matrix $\hat{\rho}_\mathrm{S}(t_0)$ of 
an isolated system (not coupled to the probes), constructed from the eigenstates of the Hamiltonian $\hat{H}_\mathrm{S}$. 
The open boundary terms are switched on over a short time interval of $5$~fs and maintained throughout the simulation. The 
current through the dot is then calculated as a bond current between the dot and the adjacent site.~\cite{bond} 
The typical parameters of the MPB setup used in our simulations, unless specified otherwise, are $N_\mathrm{L/R}=90$ and 
$N_\mathrm{d}=20$. We have tested that further increasing the size of the system does not lead to significant difference in the \textit{I-V} 
characteristics. In order to have one free parameter instead of two, we use the condition $\Delta$=$\Gamma/2$, which has been discussed 
in detail in Ref.~[\onlinecite{Todorov1}], and $\Gamma=0.35$~eV in our simulations.

\section{Results}\label{results}

\subsection{Non-interacting case}\label{nonint}

\begin{figure}[ht!]
\begin{center}\includegraphics[width=0.97\linewidth,clip=true]{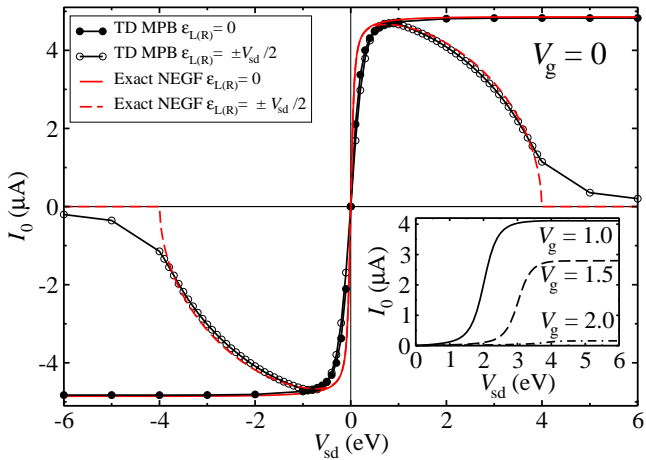}\end{center}
\caption{(Color online) Current through the quantum dot, $I_0$, as a function of the source-drain voltage, $V_\mathrm{sd}$, for zero 
gate voltage ($V_\mathrm{g}$=$0$), calculated using the both exact NEGF and the TD MPB method, and for two configurations of the leads: 
$\varepsilon_\mathrm{L/R}$=$0$ and $\varepsilon_\mathrm{L/R}$=$\pm V_\mathrm{sd}/2$. The inset shows the \textit{I-V} 
characteristics obtained with the TD MPB approach for $\varepsilon_\mathrm{L/R}$=$0$ and three different gate voltages, 
$V_\mathrm{g}$=$1.0$, $1.5$ and $2.0$~eV. The following parameters are used: $\gamma_0$=$-1.0$~eV, $\gamma_\mathrm{c}$=$-0.1$~eV and
$\varepsilon_\mathrm{F}$=$0$. The source-drain voltage is applied symmetrically, $\mu_\mathrm{L/R}$=$\varepsilon_\mathrm{F}\pm V_\mathrm{sd}/2$. 
In order to achieve a better agreement with the exact NEGF results we use the improved MPB setup with $N_{\mathrm{L(R)}}=250$, 
$N_{\mathrm{d}}=90$ and $\Gamma=0.15$.}
\label{tests}
\end{figure}

As a test of the applicability of the TD MPB method we first examine the non-interacting case ($U=0$). For this situation, we directly compare 
the \textit{I-V} characteristics obtained from the time-dependent simulations to the ones calculated by using the standard NEGF-based 
Landauer solution, which we refer to as exact NEGF.~\cite{Haug} The comparison is presented in Fig.~\ref{tests}, where the current is plotted 
as a function of the source-drain voltage for the non-interacting level aligned with the Fermi level in the leads ($V_\mathrm{g}=0$). In the case 
of the TD MPB approach, the value for the steady-state current is obtained from the time-dependent simulation for the corresponding value of 
$V_\mathrm{sd}$ after the steady-state has been established, i.e. when the variation of the current with time becomes negligible. In the case 
of the exact NEGF method, we use the well-known analytical expression for the non-equilibrium current through a non-interacting resonant level 
coupled to two semi-infinite electrodes~\cite{Haug,Baldea}
\begin{eqnarray}
  I_\mathrm{EN}=\frac{2e}{h}\int\,&dE&\,\frac{\Gamma_{\mathrm{L}}(E)\Gamma_{\mathrm{R}}(E)}
{\left[E-V_\mathrm{g}-\Lambda(E)\right]^2+\left[\Gamma(E)/2\right]^2}\times\nonumber\\
&\times&\left[f_\mathrm{L}(E)-f_\mathrm{R}(E)\right].\label{exact_negf}
\end{eqnarray}
Here $\Lambda(E)=\Lambda_\mathrm{L}(E)+\Lambda_\mathrm{R}(E)$ and $\Gamma(E)=\Gamma_\mathrm{L}(E)+\Gamma_\mathrm{R}(E)$ 
represent, respectively, the real and imaginary part of the total self-energy due to the presence of electrodes,  with 
$\Lambda_\mathrm{L(R)}$ and $\Gamma_\mathrm{L(R)}$ given by
\begin{eqnarray}
 \Lambda_\mathrm{L(R)}(E)&=&\frac{\gamma_\mathrm{c}^2}{2\gamma_0^2}E_\mathrm{L(R)},\\
 \Gamma_\mathrm{L(R)}(E)&=&\frac{\gamma_\mathrm{c}^2}{\gamma_0^2}\theta(2\gamma_0-|E_\mathrm{L(R)}|)\sqrt{4\gamma_0^2-E_\mathrm{L(R)}^2}\:,
\end{eqnarray}
where $E_\mathrm{\alpha}\equiv E-\varepsilon_\mathrm{\alpha}$, $\varepsilon_\mathrm{\alpha}$ being the on-site energy in the lead ($\alpha=$L, R). 

We consider  two possible limits for the on-site energies in the electrodes: (\textit{i}) the highly conducting regime with 
$\varepsilon_\mathrm{\alpha}$=$0$ for all atoms in $\alpha=$L, R and (\textit{ii}) the weakly conducting regime for which the on-site 
energies in $\mathrm{L(R)}$ are shifted in accordance with the respective electrochemical potential, 
$\varepsilon_\mathrm{L(R)}$=$\pm V_\mathrm{sd}/2$. As expected, the difference between the \textit{I-V} curves calculated in these two 
limits becomes significant at large bias, since the transmission in case ($ii$) rapidly drops to zero once the bias voltage exceeds the bandwidth 
of the leads ($4|\gamma_0|$). This high-bias NDC effect, stemming entirely from the finite electrode band-with, is a well-understood feature 
of steady-state transport in low-dimensional yet uncorrelated electron systems.~\cite{Baldea} We also note that the low-bias agreement 
between the two transport limits can, in principle, be extended to arbitrarily high biases $V_\mathrm{sd}$ by increasing 
$\gamma_0>V_\mathrm{sd}/4$.~\footnote{We have established that if $|\gamma_0|$ is increased from $1$~eV to $3.88$~eV, the result for 
the current, obtained using two limits for the on-site energies of the leads, differ by at most $3\%$ for $V_\mathrm{sd}$=$4$~eV and for 
$V_\mathrm{g}$ between $0$ and $1$~eV.}

An encouraging result is that for both the transport limits the TD MPB method reconstructs rather well the exact NEGF \textit{I-V}. The 
agreement is particularly good in the highly conducting limit. The smearing of the abrupt \textit{I-V} features at low bias and again the 
NDC drop at $V_\mathrm{sd}\lesssim4\gamma_0$ for the weakly conducting limit are inherent to the TD MPB method.~\cite{Todorov1} 
These are due to the explicit dephasing factor, which simplifies the equation of motion for the density matrix by eliminating temporal 
non-localities of the injection.  

In order to eliminate the drop in the current at large bias voltages and to focus on the electron interaction at the quantum dot, we will use 
the $\varepsilon_\mathrm{L(R)}$=$0$ limit in all the further calculations presented. In this case, the saturation current at high voltages 
is entirely determined by the position of the resonant level, set by the gate voltage $V_\mathrm{g}$ (see the inset of Fig~\ref{tests}), 
relatively to the electrodes band center. As the resonant the level approaches the band-edge of the leads ($V_g \lesssim 2\gamma_0$), 
the saturation current decreases. In Section~\ref{ss} we will recognize the contribution of the latter effect to the drop in the current as 
a function of the source-drain voltage. 

\subsection{Interacting case}\label{int}

\subsubsection{Time-dependent transport}\label{td}

While in the non-interacting case the TD current through the dot always reaches the steady-state, in the case when electron-electron 
interaction is considered this is not guaranteed. In fact for certain values of the source-drain voltage, for which the charge density of 
the dot approaches unity, the system is driven into a dynamical state, where current, density and on-site potential oscillate~\cite{Kurth} 
without ever reaching a steady-state. 
\begin{figure}[ht]
\begin{center}\includegraphics[width=0.97\linewidth,clip=true]{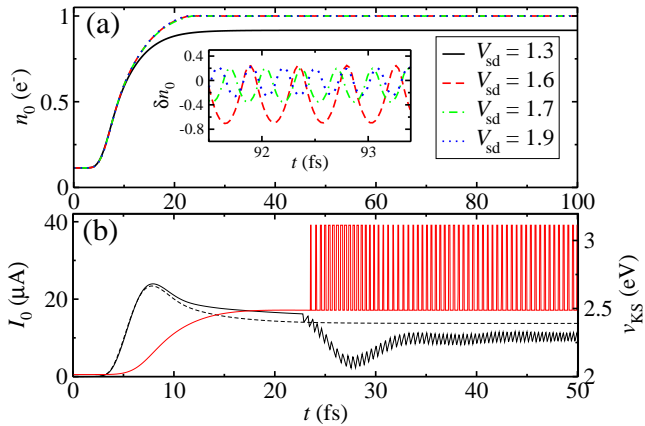}\end{center}
\caption{(Color online) Real-time evolution of the quantum dot: (a) Charge density of the dot ($n_0$) for four different values of the source-drain voltage, $V_\mathrm{sd}$=$1.3$, $1.6$, $1.7$, and $1.9$~eV. The inset shows the fluctuation of the density around unity, $\delta n_0$, defined as $\delta n_0=(n_0-1)\times 10^{3}$. (b) Current through the dot, $I_0$, for two values of $V_\mathrm{sd}$: $V_\mathrm{sd}$=$1.6$~eV (black solid line), which corresponds to the oscillating regime, and $V_\mathrm{sd}$=$1.3$~eV (black dashed line) where no oscillations are observed. Note that the corresponding Kohn-Sham potential ($v_\mathrm{KS}$) [red solid line] is also in the oscillating regime ($V_\mathrm{sd}$=$1.6$~eV). The following parameters are used: $\gamma_0$=$-1.5$~eV, $\gamma_\mathrm{c}$=$-0.3$~eV, $\varepsilon_\mathrm{F}$=$1.5$~eV, $U$=$2.0$~eV, $\varepsilon_\mathrm{L(R)}$=$0$ is taken as a reference of energy. The source-drain voltage is applied asymmetrically ($\mu_\mathrm{L}$=$\varepsilon_\mathrm{F}+V_\mathrm{sd}$, $\mu_\mathrm{R}$=$\varepsilon_\mathrm{F}$).}
\label{cb}
\end{figure}

The question we address here is whether such dynamical state can be captured by the MPB method. The results of our calculations are 
shown in Fig.~\ref{cb}. For all values of the source-drain voltage below a critical value  $V^\mathrm{cr}_\mathrm{sd}$ a steady-state is 
achieved. However, for source-drain voltages above $V^\mathrm{cr}_\mathrm{sd}$, oscillations indeed develop in all transport-related 
quantities. As shown in Fig.~\ref{cb} for this range of $V_\mathrm{sd}$ the density quickly reaches a critical value of $n_0=1$. 
At the same time the first jump of the on-site potential occurs, followed by a series of almost rectangular pulses [see Fig.~\ref{cb}(b)]. Due to 
the derivative discontinuity at $n_0=1$, the on-site potential reaches an oscillating regime, abruptly alternating in time between two values, 
one just below and the other  just above the discontinuity. This translates into oscillations of the charge density around $n_0=1$ [see the inset 
in see Fig.~\ref{cb}(a)] and also into oscillations in the current [Fig.~\ref{cb}(b)].  

Below, we elaborate on the dynamical features observed for different values of $V_\mathrm{sd}$. The height of the pulses in $v_\mathrm{KS}(t)$ 
is equal to the height of the jump of $v_\mathrm{KS}[n_0]$ at the derivative discontinuity and it is mainly governed by the value of the charging 
energy $U$. The width of the pulses increases with increasing $V_\mathrm{sd}$. This essentially means that for larger $V_\mathrm{sd}$ the 
system tends to stay longer in the state with a larger on-site potential, corresponding to the density above $1$. Further increasing $V_\mathrm{sd}$ 
will finally lead to a steady-state. The exact value of the threshold voltage, $V^\mathrm{cr}_\mathrm{sd}$, is difficult to determine since the 
on-site potential changes with time. From simple considerations, however, we established that $V^\mathrm{cr}_\mathrm{sd}\ge v_\mathrm{KS}[\bar{n}]$, 
where  $\bar{n}$ is a value of the charge density just below $1$. For the set of parameters used here $V^\mathrm{cr}_\mathrm{sd}\approx 1.5$~eV.

As discussed by Kurth \textit{et al.}, the dynamical state of the quantum dot described above is a manifestation of dynamical Coulomb 
blockade. By applying a large enough source-drain voltage the dot can be charged. However, when the charge reaches the critical value 
$n_0=1$, the on-site potential immediately increases by an amount, determined by Coulomb repulsion $U$, thus preventing further 
charging. This essentially corresponds to the CB regime. In addition, the time-dependent simulations reveal that in this regime the quantum 
dot is alternating between two states, separated by an energy barrier determined by $U$. These two states correspond to the fluctuation 
of the charge on the dot around $n_0=1$, which originates from the fact that the ABALDA potential has a derivative discontinuity at $n_0=1$ 
but it is a smoothly varying function of $n_0$ away from this occupation. 

It follows from the discussion that the dynamics of the quantum dot in the CB regime, calculated with the TD MPB method, is in a good 
agreement with the results reported in Ref.~[\onlinecite{Kurth}] both qualitatively and quantitatively. We have established numerically that 
the two different methods reproduce practically identical dynamical trajectories for all the observables in the long-time limit in the case of 
an interacting system. The remaining differences are limited to the early stage of the time-evolution. A characteristic feature of the on-site 
potential of the dot, observed in Ref.~[\onlinecite{Kurth}], is a transition period just after the start of the oscillations, where the series of 
rectangular pulses in the time-dependent $v_\mathrm{KS}$ is preceded by a larger pulse whose width increases with $V_\mathrm{sd}$. 
This characteristic transient pulse is not present in our calculations (see Fig.~\ref{cb}). 

\begin{figure}[ht]
 \begin{center}\includegraphics[width=0.97\linewidth,clip=true]{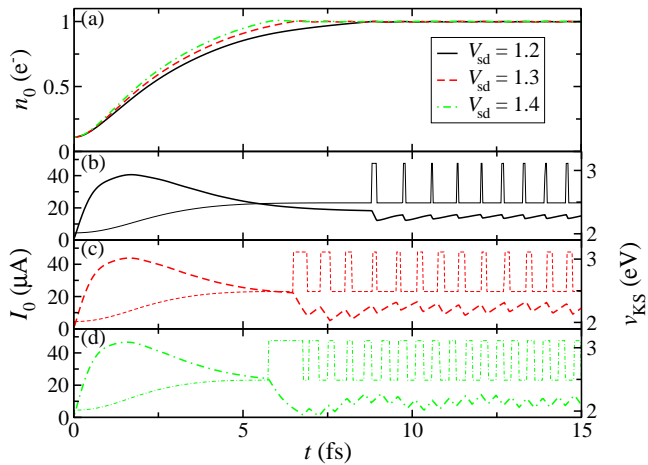}\end{center}
 \caption{(Color online) Real-time evolution of the quantum dot in the closed-boundary setup: 
(a) Charge density of the dot ($n_0$) for three different values of the 
source-drain voltage, $V_\mathrm{sd}$=$1.2$, $1.3$, and $1.4$~eV. Current through the dot, $I_0$, [thick lines] and the 
corresponding Kohn-Sham potential, $v_\mathrm{KS}$, [thin lines] 
for (b) $V_\mathrm{sd}$=$1.2$~eV, (c) $V_\mathrm{sd}$=$1.3$~eV,  and (d) $V_\mathrm{sd}$=$1.4$~eV. 
The parameters are the same as in Fig.~\ref{cb}. 
The source-drain voltage is applied as a rigid shift of the on-site energies in the left lead.}
\label{cb_closed}
\end{figure}

In order to establish to what extent the transient pulse is determined by the initial conditions, we performed TD simulations for the same 
system as shown in Fig.~\ref{qd} but without attaching the external probes, i.e. for a closed-boundary finite system. Instead, we applied 
the source-drain voltage as a rigid shift of the on-site energies in the left lead, i.e. a term  
$V_\mathrm{sd}\sum_{\scriptsize i,\sigma}\hat{c}^{\sigma\dagger}_{i\alpha} \,\hat{c}^{\sigma}_{i\alpha}$ 
has been added to the Hamiltonian $\hat{H}_\mathrm{\alpha}$ for $\alpha=\mathrm{L}$ [see Eq.~(\ref{eq:1a})] at the start of the TD simulation. 
We used longer leads ($N_\mathrm{L/R}$=$220$) and limited 
the time of the simulations to  $100$~fs, which is sufficient to observe the time propagation before the reflections from the finite boundaries 
start to affect the dynamics. The time-dependence of the 
charge density, current and on-site potential, obtained from the closed-boundary simulation, is presented in Fig.~\ref{cb_closed}. 
In contrast to our open-boundary simulations, we indeed observed qualitatively the same transient regime 
as in Ref.~[\onlinecite{Kurth}]. This is mainly characterized by an earlier onset of the CB oscillations for larger source-drain voltages and by 
the increase of the width of the first pulse in the time-dependence of the Kohn-Sham potential with increasing $V_\mathrm{sd}$.

\subsubsection{Steady-state transport}\label{ss}

In the previous section we demonstrated that, within a certain range of parameters, the derivative discontinuity prevents the quantum dot 
to evolve towards the steady-state. Outside this range, however, a steady-state is achievable. Here we determine the steady-state current 
through the dot for various gate voltages and map out the corresponding \textit{I-V} curves. For situations, where the dot is trapped in 
oscillations, we take as steady-state current its time-average in the long-time limit. 

\begin{figure}
 \begin{center}\includegraphics[width=0.97\linewidth,clip=true]{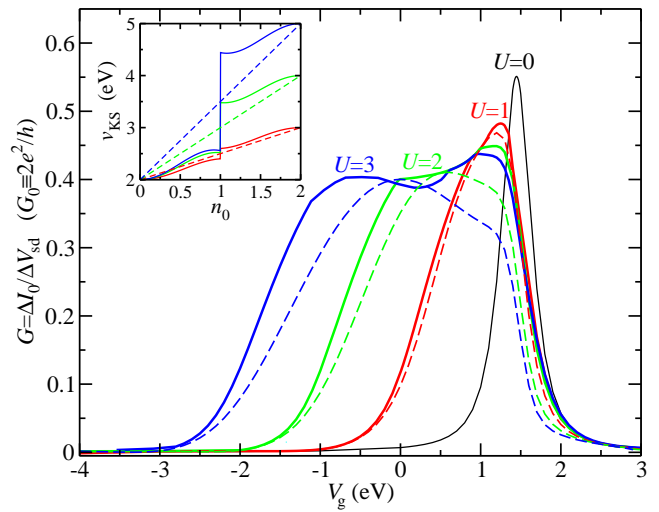}\end{center}
 \caption{(Color online) Differential conductance of the dot as a function of the gate voltage ($V_\mathrm{g}$) for the Kohn-Sham 
 potential, $v_\mathrm{KS}$, [thick solid lines] and for the Hartree potential, $v_\mathrm{H}$, [dashed lines] with $U$=$1$, $2$, and $3$~eV, 
 and for the non-interacting case (thin solid line). The inset shows a comparison between the density-dependence of $v_\mathrm{KS}$ 
 (solid lines) and $v_\mathrm{H}$ (dashed lines) for the same values of $U$. Parameters are the same as those of Fig.~\ref{cb} and 
 $V_\mathrm{sd}=0.01$~eV.}
 \label{iv}
 \end{figure}

The linear response conductance as a function of $V_\mathrm{g}$ is depicted in Fig.~\ref{iv}. This is calculated as the finite-difference 
ratio $\Delta I_0/\Delta V_\mathrm{sd}$ close to zero bias (for a very low but finite bias $\Delta V_\mathrm{sd}=0.01$ eV) and represents 
an approximation to the zero-bias differential conductance. In the non-interacting case, the conductance is composed of a single peak 
centered around $V_\mathrm{g}$=$1.5$~eV, which corresponds to the Fermi level of the leads. This is expected from the steady-state 
picture of transport through a non-interacting resonant level. In principle the width of the resonance peak is given by the dot-lead hopping 
integral $\gamma_\mathrm{c}$. In our TD MPB calculations, however, there is an additional resonance broadening factor ($\tau_\Delta$) 
related to the dephasing condition in the equations of motion. Its corresponding energy unit, $\Delta=\hbar/\tau_\Delta$, can 
be associated to a fictitious temperature, smearing the electronic energy distributions in the leads.~\cite{Todorov1} As a result, a suppression 
of the transmission resonance proportional to $1/\Delta$ is also expected. This is the reason of why the amplitude of non-interacting 
resonance conductance in Fig.~\ref{iv} is below one quantum of conductance, $G_0=2e^2/h$. 

In the interacting case the Anderson impurity model predicts two distinct Coulomb peaks~\cite{Hanson} in the conductance as a function of the 
gate voltage~\cite{Bruus}. These are manifestation of charge quantization at the dot and correspond to each of the two integer electron number 
states, in which the dot is inhabited by one or two electrons, respectively. Although the ABALDA potential succeeds in describing some important 
properties of strongly correlated systems,~\cite{Akin} due to the presence of the derivative discontinuity, it is a single-particle potential and, 
as such, cannot describe fully these charge states. As a result, the gate-voltage dependence of the conductance, calculated using the full 
discontinuous effective potential ($v_\mathrm{KS}$), does not show two distinct peaks. However, it presents a structure, bearing the signature 
of two broadened and overlapping peaks (see Fig.~\ref{iv}). The distance between these quasi-peaks increases with increasing $U$ and 
corresponds to the value of the jump of the on-site potential $v_\mathrm{KS}[n_0]$ at the derivative discontinuity. In the case of the Hartree 
approximation, the two-peak structure is less pronounced and the two resonances merge into an asymmetric plateau. The width of this plateau 
is also proportional to $U$.  

It should be mentioned that for the TD calculations with $v_\mathrm{KS}$ and for values of $V_\mathrm{g}$ between the position of the 
$U=0$ resonance level $V_\mathrm{res}\equiv \varepsilon_F$ and $V_\mathrm{res}-U$ (roughly corresponding to the region between the 
two quasi-peaks) no steady-state is achieved. Hence, the conductance curves in this region of $V_\mathrm{g}$ carry some degree of 
arbitrariness, associated with the interpretation of the average TD current. In fact, for those gate voltages driving a charge density at the 
dot close to unity, even the calculation of the ground-state is problematic from a numerical viewpoint, because of the derivative discontinuity. 
In such cases we used the following iterative procedure. Let $V_\mathrm{g}^{0}$ be the value of the gate voltage, for which the ground-state 
(initial) density is calculated self-consistently, while $V_\mathrm{g}^{0}+\delta V_\mathrm{g}$ is the value of the gate voltage for which the 
self-consistent calculation does not converge. In this case, the final density, obtained at the end of the time-dependent simulation with 
$V_\mathrm{g}$=$V_\mathrm{g}^{0}$, is taken as initial density for the simulation with $V_\mathrm{g}$=$V_\mathrm{g}^{0}+\delta V_\mathrm{g}$.

\begin{figure}
\begin{center}\includegraphics[width=0.97\linewidth,clip=true]{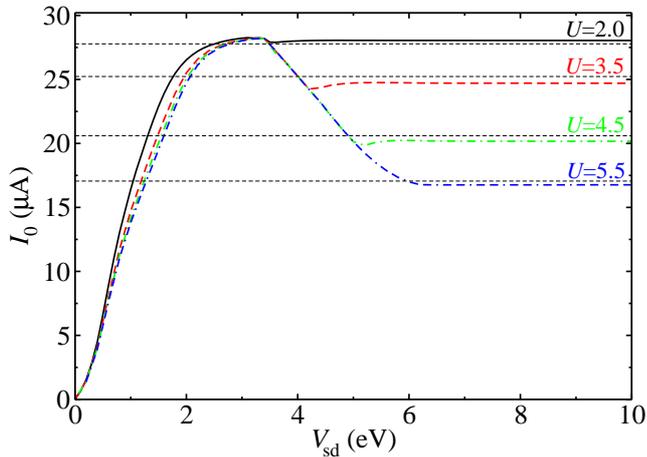}\end{center}
 \caption{(Color online) Current through the dot, $I_0$, as a function of the source-drain voltage, $V_\mathrm{sd}$,  
 for $v_\mathrm{KS}$ and different values of $U$. The horizontal dashed lines represent the corresponding saturation 
 currents $I_\mathrm{S}$ (see text for the exact definition). The following parameters are used: $\gamma_0$=$-3.88$~eV, 
 $\gamma_\mathrm{c}$=$-0.5$~eV, $\varepsilon_\mathrm{F}$=$1.5$~eV, $V_\mathrm{g}$=$2.0$~eV. 
}
\label{iv1}
 \end{figure}

In the same way, from the time-averages in the long time-limit, we map out the \textit{I-V} characteristics of the interacting dot ($v_\mathrm{KS}$) 
at a given $V_\mathrm{g}$ (see Fig.~\ref{iv1}). A remarkable feature of the \textit{I-V} curves is the drop of the current (NDC) at large source-drain 
voltages, which is almost negligible for small $U$ but increases with increasing $U$. 

\begin{figure}
 \begin{center}\includegraphics[width=0.97\linewidth,clip=true]{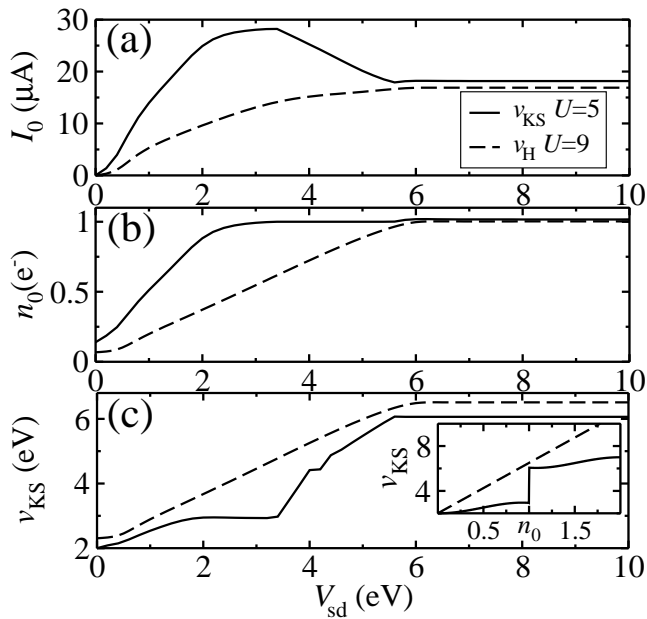}\end{center}
 \caption{Current (a), density (b) and on-site potential (c) of the dot as a function of the source-drain voltage, $V_\mathrm{sd}$,
for $v_\mathrm{KS}$ with $U=5$~eV and for  $v_\mathrm{H}$ with $U=9$~eV. The inset shows $v_\mathrm{KS}$  
and $v_\mathrm{H}$ as functions of the dot density for the corresponding values of $U$. The parameters are the same as 
those in Fig.~\ref{iv1}.}
 \label{iv2}
 \end{figure}

For all values of $U$ the current initially increases with increasing $V_\mathrm{sd}$ as the dot is charging. It then reaches its 
maximum value as the charge density approaches $n_0=1$. This point corresponds to a threshold source-drain voltage 
$V^\mathrm{cr}_\mathrm{sd}$, which is roughly the same for all values of $U$. Beyond $V^\mathrm{cr}_\mathrm{sd}$, the 
system is driven into a dynamical state (where the steady-state current is calculated by averaging out the oscillations). In the 
limit of very large $V_\mathrm{sd}$, the dot recovers its long-time tendency to a steady state and the average current saturates. 
At saturation and beyond the dot occupation is above $1$ and the on-site potential assumes a value above the discontinuity. 
Hence, the on-site energy at the dot is proportional to the the jump of the $v_\mathrm{KS}$ at $n_0=1$, i.e. it is proportional to $U$. 

As discussed in Section~\ref{nonint} for the non-interacting case, the saturation current decreases with increasing the dot on-site potential, 
because of the finite bandwidth of the electrodes. For the same reason here the drop of the current becomes larger when $U$ 
increases. In fact a large $U$ corresponds to a large value of the steady-state on-site potential, which then approaches the electrodes'
band-edge. In order to confirm this conjecture, we compare the saturation current $I_\mathrm{S}$ calculated at finite $U$, with that
for $U$=$0$ and $V_\mathrm{g}$ equal to the steady-state on-site potential corresponding to that obtained at the same $U$. 
Indeed $I_\mathrm{S}$ matches quite well the value of the current obtained at large source-drain voltages in the \textit{I-V} characteristics 
of the interacting dot (see horizontal dashed lines next to each curve in Fig.~\ref{iv1}). This argument can obviously be reversed, i.e. 
the NDC cannot be observed, if the variation of the on-site potential at the derivative discontinuity, determined by $U$, is much smaller 
than the electrodes' bandwidth. For instance, for the same set of parameters used before for the dot+electrodes system, 
such NDC-free situation is found for $U=2$~eV ($U\ll4|\gamma_0|$ for $\gamma_0=3.88$~eV). In this case the drop of the current 
above $V^\mathrm{cr}_\mathrm{sd}$ is practically negligible. 

Importantly, the NDC displayed in Fig.~\ref{iv1} is not found in \textit{I-V}'s calculated within the Hartree approximation, even for 
large values of $U$ (see Fig.~\ref{iv2}). When comparing calculations at the Hartree level with those performed with the complete
Kohn-Sham potential we intentionally use different $U$. These are selected in such a way that the value of the potential at
$n_0=1$ is identical in the two calculations [see the inset in Fig.~\ref{iv2}(c)], i.e. in such a way that the two calculations give the same
saturation current. At variance with the complete Kohn-Sham case, in the Hartree only problem the current, as well as the density and 
the on-site potential, monotonically increase with $V_\mathrm{sd}$ until the saturation is reached. Based on these numerical results 
we can argue that the self-interaction-free shape of the on-site potential $v_\mathrm{KS}$ at the dot is a necessary condition for the 
occurrence of the NDC in the \textit{I-V}. The shallow increase of the on-site potential with the charging, produced by the opening of 
the bias window, keeps the resonant level away from the electrodes band edge and allows the current to rise. Once the on-site charge 
exceeds $n_0=1$ and the resonant level energy shoots up towards the band-edge, the currents drops. The averaged dynamical 
current monotonically approaches its saturation value corresponding to a steady-state solution.  

\section{Conclusions}\label{concl}

We have investigated the electronic transport through a strongly-correlated quantum dot by using a recently proposed multiple-probe 
battery method for time-dependent simulations of open systems. Our aim was two-fold. Firstly, we wanted to assess the outcomes 
of a TD transport scheme conceptually different from what used so far in literature, for a problem involving strong electron correlation
as in Coulomb blockade. Clearly our MPB-based simulations agree well with previous findings.~\cite{Kurth} In particular we have 
demonstrated self-sustained oscillations in the current, density and effective on-site potential, originating from the derivative discontinuity 
of the approximate exchange-correlation potential used. 

As a further aspect we have addressed the question of whether the peculiar dynamics obtained from the time-dependent simulations can 
be related to the more accessible steady-state picture of transport. In particular, we have shown the presence of Coulomb peaks in the 
linear response differential conductance and extracted the TD version of \textit{I-V} characteristics, based on the time-averaged current 
through the dot in the long-time limit. The resulting \textit{I-V} curves, at a critical voltage, exhibit a drop in the average current through 
the dot. This drop corresponds to the range of parameters where no steady state is found and the dot is in the oscillatory Coulomb 
blockade state. Such an NDC is however present only when the calculation is performed at a DFT level in which the potential includes the
derivative discontinuity at unitary occupation.

\begin{acknowledgments}
 
We are grateful to A.~Hurley and I.~Rungger for careful reading of the manuscript. We thank T.~N.~Todorov for very helpful 
discussions. This work is sponsored by Science Foundation of Ireland (Grant No. 07/IN.1/I945). Computational resources 
have been provided  by the Trinity Center for High Performance Computing.

\end{acknowledgments}

\end{document}